\documentclass[final,5p,times]{elsarticle}
\usepackage{multirow}
\journal{Scripta Materialia}
\usepackage{hyperref}
\usepackage{dblfloatfix}
\usepackage[utf8]{inputenc}

\usepackage{amssymb}
\usepackage{amsmath}
\usepackage{upgreek}
\usepackage{algorithm}
\usepackage{array}
\usepackage{bbold}

\usepackage{stackengine}

\usepackage{graphicx}%Einbinden von Bildern
\usepackage{caption}
\usepackage{subcaption}
\usepackage[usenames]{color}%farbiger Text
\usepackage[labelfont=bf,textfont=normalsize]{caption}
\usepackage[labelfont=bf,textfont=small]{subcaption}

\usepackage{here} %Balduin: ergänzt für Anordnung Grafiken
\usepackage{multicol}
\usepackage{wrapfig}
\usepackage{float}
\usepackage{tabularx}
\usepackage[export]{adjustbox}
\usepackage{pgf}
\usepackage{stackengine}
\usepackage{textcomp}
\usepackage{tablefootnote}

\usepackage{soul,xcolor} %For Cross out

\usepackage{jabbrv}

\usepackage{comment} %Balduin: ergänzt für Kommentieren
\usepackage{outlines}
\usepackage{booktabs}

\usepackage{color}
\usepackage[normalem]{ulem}
\newcommand{\Delete} [1]{\bgroup\noindent\textcolor{red}{\xout{#1}}\egroup\ignorespacesafterend}
\newcommand{\Insert} [1]{\bgroup\noindent\textcolor{blue}{#1}\egroup\ignorespacesafterend}
\newcommand{\frage}[1]{{#1}}
\newcommand{\budgetremark}[1]{\frage{\color{brown} [budget: #1]}}
\renewcommand{\budgetremark}[1]{}

\setstcolor{red} %Cross Out Color

\date{06. December 2022}

\begin{document}

%\twocolumn[{\begin{frontmatter} %two column
\begin{frontmatter} %one column



\title{Characterization of Lomer junctions based on the Lomer arm length distribution in dislocation networks}
%


% \author{}
% 
% \address{}

%\author[InstitutOfAB]{A B\corref{cor1}}

%\fntext[label2]{These authors contributed equally to this work.}
%\ead{a.b@kit.edu}
%% \ead[url]{http://www.InstitutOfAB.kit.edu}
%\author[IAM,HS]{Balduin Katzer\fnref{label2}}
%\author[IAM,HS]{Kolja Zoller\fnref{label2}}

\author[IAM,HS]{Balduin Katzer}
\author[IAM,HS]{Kolja Zoller}
\author[IAM]{Julia Bermuth}
\author[IAM]{Daniel Weygand}
\author[IAM,HS]{Katrin Schulz\corref{cor1}}

\cortext[cor1]{Corresponding author.}

\address[IAM]{Karlsruhe Institute of Technology (KIT), Institute for Applied Materials (IAM),%\\
              Kaiserstr. 12, 
              76131 Karlsruhe, Germany}
\address[HS]{Karlsruhe University of Applied Sciences, Moltkestr. 30, 76133, Karlsruhe, Germany}
\ead{katrin.schulz@kit.edu}

%\ead[url]{https://www.iam.kit.edu/cms/english/index.php}
             
%\author[institute]{name}
%\address[institute]{name of institute,
%              Karlsruhe Institute of Technology (KIT),\\
%              Kaiserstr. 12, 
%              76131 Karlsruhe, Germany}

\begin{abstract}
%%% Text of abstract
%%% ABSTRACT

%TC:ignore
During the plastic deformation of crystalline materials, 3d dislocation networks form based on dislocation junctions.
Particularly, immobile Lomer junctions are essential for the stability of dislocation networks.
However, the formed Lomer junctions can unzip and dissolve again, if the linked mobile dislocations of the Lomer junction - the Lomer arms - experience sufficiently high resolved shear stresses.
To generate a better understanding of the dislocation network stability and to pave the way to a general stability criterion of dislocation networks, we investigate the Lomer arm length distribution in dislocation networks by analyzing discrete dislocation dynamics simulation data of tensile-tested aluminum single crystals.
We show that an exponential distribution fits best to the Lomer arm length distribution in the systems considered, which is independent of the crystal orientation.
The influence of the slip system activity on the Lomer arm length distribution is discussed.

\begin{keyword}
Dislocation dynamics \sep
Network stability \sep
Lomer junction \sep
Single crystal \sep
Plastic deformation \sep
DDD \sep
CDD
\end{keyword}
%TC:endignore

\end{abstract}
%}] %two column

%\begin{keyword}
%%% keywords here, in the form: keyword \sep keyword
%%% PACS codes here, in the form: \PACS code \sep code
%%% MSC codes here, in the form: \MSC code \sep code
%%% or \MSC[2008] code \sep code (2000 is the default)
%\end{keyword}

\end{frontmatter} %one column


%\tableofcontents
%\listoffigures
%\listoftables
%\newpage

%%% FULL-TEXT

Dislocation network structures evolve in crystalline materials during plastic deformation.
Ongoing plastic deformation can be observed by tracking the motion of mobile dislocations with respect to the resolved shear stress on individual slip systems.
However, dislocation networks consist of mobile and immobile dislocations.
The motion of dislocations is hindered by mutual interactions between the dislocations, which largely leads to strain hardening~\cite{seeger1954,saada_sur_1960,Schoeck1972}.
Thus, the characterization of dislocation networks has to incorporate two competing mechanisms: The formation of stable network structures and the mobility of dislocations in network structures \cite{madec_role_2003,weygand_study_2005}.

In fcc crystals, the interaction of dislocations on two different slip systems may lead to junctions.
Junctions are thus "shared" by at least two dislocations on different slip systems.
Reactions can generate new mobile dislocations on other slip systems, e.g. due to glissile interactions or represent the cross-slip mechanisms, as discussed in \cite{Stricker2015,sudmanns_dislocation_2019}.
In contrast, reactions can also lead to immobile junctions, the so-called Lomer junctions.
The Lomer junction is considered to be sessile, since it is not part of a slip system \cite{Lomer1951}.

However, immobile Lomer junctions can also dissolve during the dislocation network evolution.
Line tension models~\cite{saada_sur_1960}, simulations of quasicontinuum and molecular dynamics (MD) \cite{Rodney1999,Weinberger2011} and discrete dislocation dynamics simulations (DDD)~\cite{Weygand_2002,madec2002,Motz2009} can be used to analyze the stability or dissolution of Lomer junctions. 
The unzipping of a Lomer junction depends on its neighboring Lomer arms, which are connected to one of the ends of the Lomer junction~\cite{saada_sur_1960, Shenoy2000, Shin2001, Rodney1999}.
In the following, we refer to the mobile dislocations, which are connected to the end-nodes of Lomer junctions as ''Lomer arms''.
Lomer arms are glissile on their respective slip system.
The ability to move a Lomer arm correlates inversely with its arm length, i.e. shorter Lomer arms require higher critical shear stresses and thus stabilize the dislocation network.
In contrast, longer Lomer arms can bow-out at lower shear stresses and thus unzip the Lomer junction.

\fussy
Consequently, the Lomer arm length is important for two aspects, the stability and dissolution of the junction within the dislocation network.
In this work, we analyze the length distribution of Lomer arms by analyzing 3d DDD data of plastically deformed fcc single crystals mimicking aluminum to lay the foundations for a general stability criterion of dislocation networks.

We analyze data of 24 DDD simulations of tensile test for the crystal orientations $\langle 100 \rangle$, $\langle 110 \rangle$, $\langle 111 \rangle$, and $\langle 123 \rangle$. For each orientation, six simulations with different relaxed initial dislocation microstructures are performed.
The used DDD framework is described in~\cite{Weygand_2001,Weygand_2002}.
The samples analyzed are tensile-tested fcc single crystal with a volume of $5\times5\times5 \; $\textmu$\mathrm{m^3}$ at a strain rate of 5000~$\mathrm{s^{-1}}$.
The cross-slip mechanism is included in the simulations~\cite{Weygand_2002}. 

\begin{table}[t]
    \setlength\tabcolsep{4pt}
    \centering
    \caption{Fraction of Lomer arms and of dislocation reactions\protect\footnotemark adjacent to Lomer junctions according to number in the initial ($\varepsilon_{tot}~=~0.0~\%$) and in the final load step ($\varepsilon_{tot}~=~0.3~\%$) of the simulations for different crystal orientations.}
    \begin{tabularx}{\linewidth}{c p{32mm} cccc}
    \toprule[1pt]\midrule[0.3pt]
        $\varepsilon_{tot}$ & \parbox{30mm}{Fraction of links next to Lomer junctions} & $\langle 100 \rangle$ & $\langle 110 \rangle$ & $\langle 111 \rangle$ & $\langle 123 \rangle$ \\
        \hline
        \multirow{2}{*}{0.0$\%$} & Lomer arm [$\%$]& 45.75 & 42.20 & 42.81 & 34.51 \\
             & Reaction [$\%$] & 54.25 & 57.80 & 57.19 & 65.49 \\
        \hline
        \multirow{2}{*}{0.3$\%$} & Lomer arm [$\%$] & 32.62 & 31.00 & 26.52 & 25.15 \\
            & Reaction [$\%$] & 67.38 & 69.00 & 73.48 & 74.85 \\
    \midrule[0.3pt]\toprule[1pt]
    \end{tabularx}
    \label{tab:fraction_Lomer_arms}
\end{table}
\begin{figure}[t]
    \centering
    \begin{subfigure}[b]{0.26\textwidth}
        \centering
        \includegraphics[height=5.1cm]{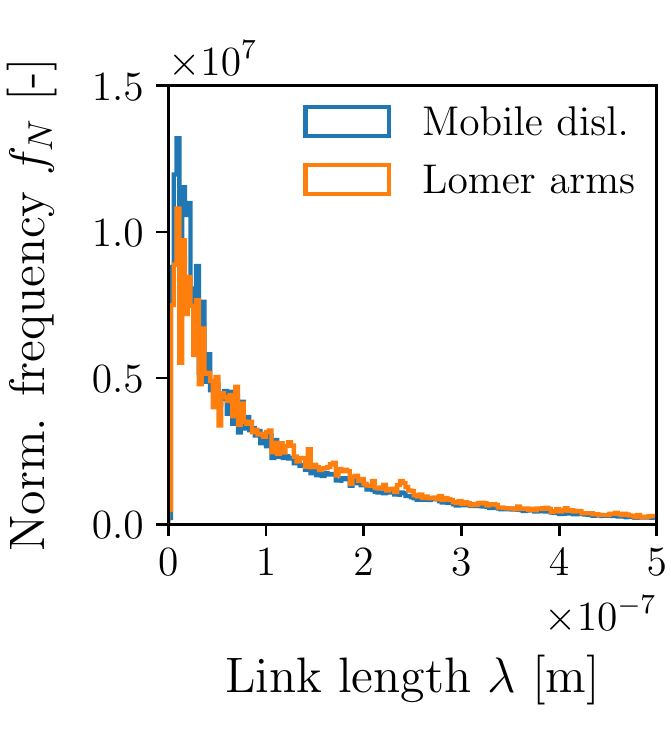}
        \llap{
        \parbox[b]{0.7cm}{(a)\\\rule{0ex}{4.55cm}
        }}
    \end{subfigure}
    \hfill
    \begin{subfigure}[b]{0.2\textwidth}
        \centering
        \includegraphics[height=5.1cm]{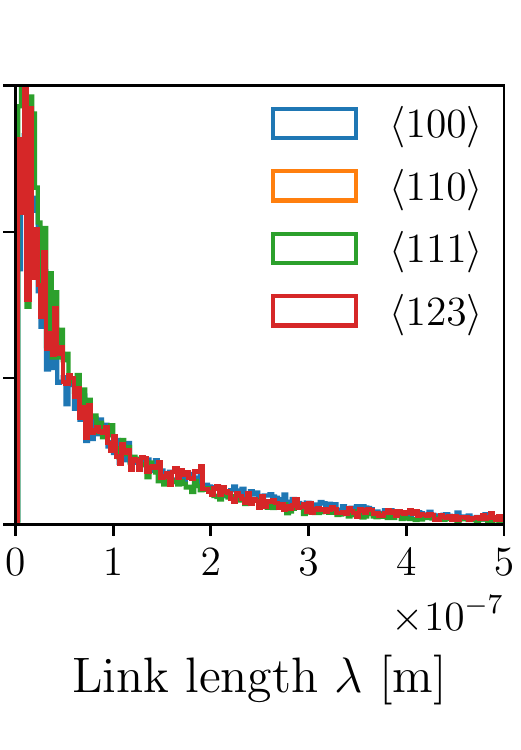}
        \llap{
        \parbox[b]{0.7cm}{(b)\\\rule{0ex}{4.55cm}
        }}
    \end{subfigure}
    \caption{Normalized link length distribution of (a) all mobile links and Lomer arms across all simulation steps of one simulation in $\langle 100 \rangle$ orientation and (b) the normalized distribution of Lomer arms for all orientations at $0.3\%$ total strain.}
    \label{fig:Lomer_arm_all_orient}
\end{figure}
We evaluate the fraction of Lomer arms in the network.
Comparing the fraction of Lomer arms for a loading of $\varepsilon_{tot}=0.3\%$ with the initial relaxed network configuration ($\varepsilon_{tot}=0.0\%$) shows a decreasing trend, see \autoref{tab:fraction_Lomer_arms}.
The dislocations involved in the Lomer reaction in turn can react again and form further reactions, which can effectively cause one of the end-nodes of the Lomer reaction to become immobile.
This is subsumed under the term ''reaction'' in \autoref{tab:fraction_Lomer_arms}.
Those Lomer junctions can then only unzip from one side.
Thus, we observe a gathering of reactions at Lomer junctions during plastic deformation for each crystal orientation considered.

\autoref{fig:Lomer_arm_all_orient}(a) shows the length distribution of all mobile dislocation links compared to the Lomer arms in~$\langle 100 \rangle$ orientation.
We observe a resembling distribution for the overall length of mobile dislocations and the Lomer arm length.
\autoref{fig:Lomer_arm_all_orient}(b) illustrates the length distribution of the Lomer arms in $\langle 100 \rangle$, $\langle 110 \rangle$, $\langle 111 \rangle$ and $\langle 123 \rangle$ orientation at a total strain of~$0.3\%$. 
The results show barely any orientation dependency.
In order to investigate and compare the Lomer arm length distribution with the general dislocation link length distribution from the literature, the following distributions are considered here, whereby from now on, the length of a Lomer arm is termed ''link length'':

\footnotetext{
The number and topology of dislocation reactions, that are different from Lomer (and Hirth) junctions, are tracked by virtual junctions, which have a zero net Burgers vector. This concept is used in this work to track the Lomer junction neighbors. For a detailed explanation we refer to \cite{Weygand_2002, Stricker_2018, 2022_Katzer_JotMaPoS}.}

Shi and Northwood \cite{shi1993} provide an analytical approach based on the considerations of Wang et al. \cite{Wang1992}, which leads to
\begin{equation}\label{eq:dis_shi}
    \phi(\lambda) = 2\rho \left( \frac{\lambda^2}{\lambda_m^4} \right) \exp{\left(-\frac{\lambda^2}{\lambda_m^2}\right)} \quad \mathrm{with} \quad L=\frac{2}{\sqrt{\pi}} \lambda_m.
\end{equation}
Thereby, the probability $\phi$ for a discrete link length $\lambda$ depends on the dislocation density $\rho$ and the link length with the highest probability $\lambda_m$, which relates to the average link length $L$.

Hu and Cocks~\cite{Hu2016} show a similar analytical solution with
\begin{equation}\label{eq:dis_hu}
    \phi(\lambda) = \frac{2\pi\lambda}{L^4} \exp{\left(-\frac{\pi\lambda^2}{L^2}\right)}.
\end{equation}
They suggest to apply the equation on individual slip systems, since they expect differing link length distributions for each slip system. 

Sills et al. \cite{Sills_2018} use a one dimensional Poisson process for the description of the link length probability and compare it to link length data obtained by DDD simulations, which results to the exponential distribution by
\begin{equation}\label{eq:dis_sills}
    \phi(\lambda) = \frac{1}{L} \exp{\left(-\frac{\lambda}{L}\right)}
\end{equation}
They suggest to replace the average link length by $L=\frac{\rho}{N}$, with the number $N$ of dislocation links per unit volume.

Shishvan and Van der Giessen~\cite{shishvan2010} investigate the different dislocation lengths of Frank-Read sources in polycrystalline materials, which resemble hindering objects such as grains or particles.
They choose a log-normal distribution, which leads to
\begin{equation}\label{eq:dis_giessen}
    \phi(\lambda) = \frac{1}{\lambda c_2 \sqrt{2 \pi}} \exp{\left(-\frac{ \left( \ln{\lambda} - c_1\right)^2}{2 c_2^2}\right)}
\end{equation}
with $c_1 = \ln{L}$, $c_2^2 = 2 \ln{\left( \frac{L}{L_{50}} \right)}$ and $L_{50}$ as the median of link lengths.

Zoller et al.~\cite{Zoller2021} use a Rayleigh distribution to approximate the link length of stabilized dislocations due to Lomer junctions, which leads to
\begin{equation}\label{eq:dis_zoller}
    \phi(\lambda) = \frac{\lambda}{\sigma_L^2} \exp{\left(-\frac{\lambda^2}{2 \sigma_L^2}\right)},
\end{equation}
where $\sigma_L$ is the expected value of the link lengths.
\begin{figure*}[t]
    \centering
    \begin{subfigure}[b]{\textwidth}
        \centering
        \includegraphics[scale=0.77,right]{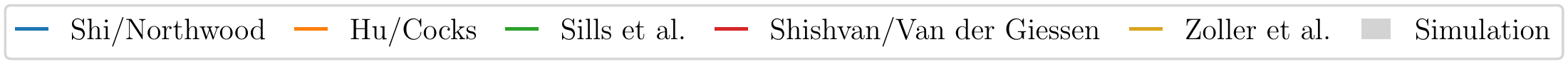}
    \end{subfigure}
    \begin{subfigure}[b]{0.26\textwidth}
        \centering
        \includegraphics[height=4.75cm]{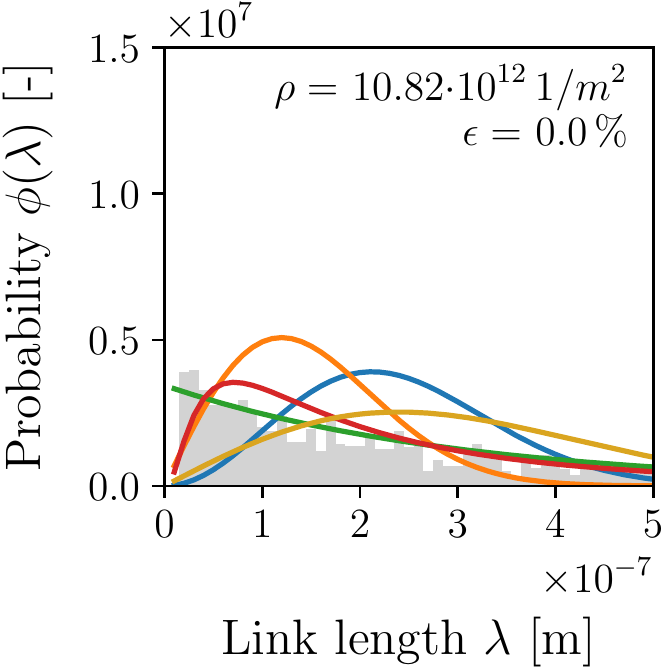}\llap{
          \parbox[b]{0.6cm}{(a)\\\rule{0ex}{4.4cm}
          }}
    \end{subfigure}
    \hfill
    \begin{subfigure}[b]{0.23\textwidth}
        \centering
        \includegraphics[height=4.75cm]{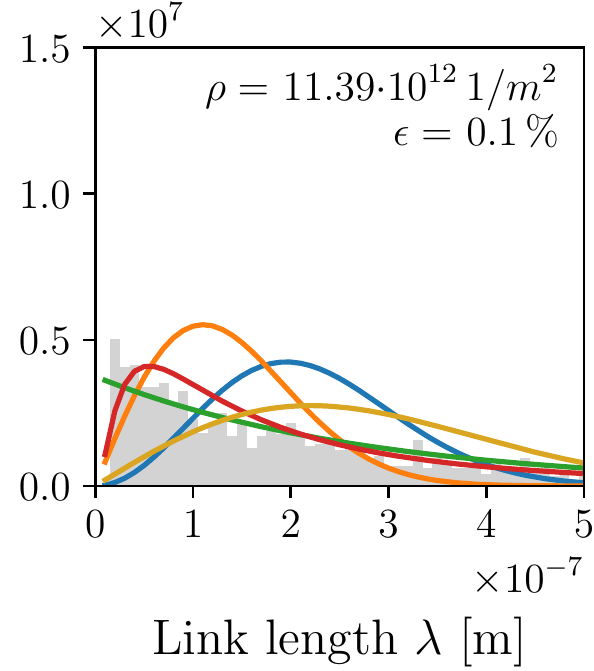}\llap{
          \parbox[b]{0.6cm}{(b)\\\rule{0ex}{4.4cm}
          }}
    \end{subfigure}
    \hfill
    \begin{subfigure}[b]{0.23\textwidth}
        \centering
        \includegraphics[height=4.75cm]{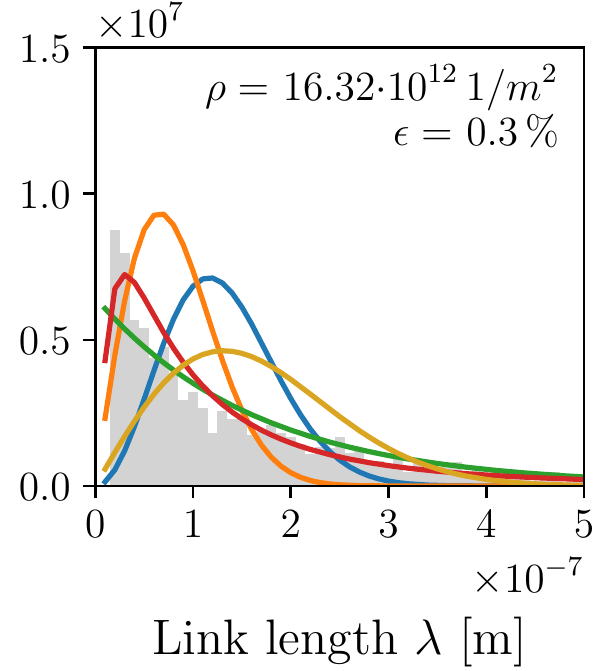}\llap{
          \parbox[b]{0.6cm}{(c)\\\rule{0ex}{4.4cm}
          }}
    \end{subfigure}
        \hfill
    \begin{subfigure}[b]{0.23\textwidth}
        \centering
        \includegraphics[height=4.75cm]{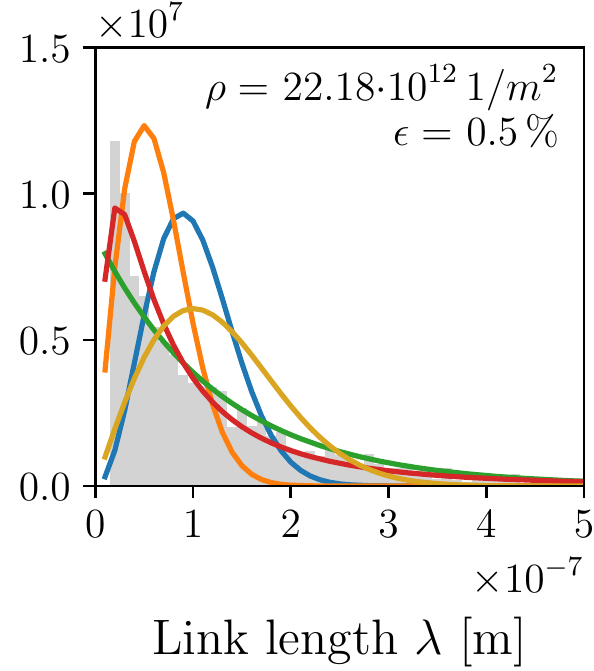}\llap{
          \parbox[b]{0.6cm}{(d)\\\rule{0ex}{4.4cm}
          }}
    \end{subfigure}
    \caption{Comparison of different link length distributions in $\langle 100 \rangle$ orientation at the initial state (a) and for (b) $0.1\%$, (c) $0.3\%$ and (d) $0.5\%$ total stain with Shi/Northwood: Analytical 1 (\autoref{eq:dis_shi}), Hu/Cocks: Analytical 2 (\autoref{eq:dis_hu}), Sills et al.: Exponential (\autoref{eq:dis_sills}), Shishvan/Van der Giessen: Log-normal (\autoref{eq:dis_giessen}) and Zoller et al.: Rayleigh (\autoref{eq:dis_zoller}. The link length distribution of the underlying data is presented in the grey bars. }
    \label{fig:distribution_compared}
\end{figure*}
The comparison of five distributions and the present data is shown in \autoref{fig:distribution_compared} for the $\langle 100 \rangle$ orientation for three different total strains, respectively.
The link length distributions for the other crystal orientations are found to show qualitatively similar distributions (cp. \autoref{fig:Lomer_arm_all_orient}(b)).

We observe that the exponential and the log-normal distribution fit best for the link length distribution.
The distribution of the link lengths shifts towards shorter lengths with increasing plastic strain and dislocation density, respectively (see \autoref{fig:distribution_compared}(a) to (d)).
The exponential distribution slightly underestimates the probability of short links, however, it applies best for longer links.
The increase of shorter links during straining is due to a densification of the dislocation network. 
\begin{table}[t]
    \centering
    \caption{Kolmogorov-Smirnov statistics for the link length distribution in $\langle 100 \rangle$, $\langle 110 \rangle$, $\langle 111 \rangle$ and $\langle 123 \rangle$ orientation of various distributions.}
    \setlength\tabcolsep{7pt}
    \begin{tabularx}{\linewidth}{{l}*{4}{c}}
        \toprule[1pt]\midrule[0.3pt]
         Distribution &  $\langle 100 \rangle$ &  $\langle 110 \rangle$ &  $\langle 111 \rangle$ &  $\langle 123 \rangle$  \\ [1ex] \hline
         Analytical 1 (\autoref{eq:dis_shi}) & 0.297 & 0.303 & 0.298 & 0.294 \\
         Analytical 2 (\autoref{eq:dis_hu}) & 0.300 & 0.290  & 0.300  & 0.299  \\
         Exponential (\autoref{eq:dis_sills}) & 0.106 & 0.107  & 0.109  & 0.102  \\
         Log-normal (\autoref{eq:dis_giessen}) & 0.116 & 0.118  & 0.122  & 0.121  \\
         Rayleigh (\autoref{eq:dis_zoller}) & 0.293 & 0.295  & 0.295  & 0.292  \\
         \midrule[0.3pt]\toprule[1pt]
    \end{tabularx}
    \label{tab:ks_stats}
\end{table}
To quantify these observations, we apply the Kolmogorov-Smirnov(ks) statistics to the distribution results.
The ks-statistic analyzes the deviation of a distribution with the data~\cite{Massey1951}.
Thereby, small values indicate a small divergence.
The ks-statistic results are listed in \autoref{tab:ks_stats} for each distribution in each crystal orientation.
The exponential and the log-normal distribution fit best in each orientation with similar ks-values.
So, we state that the qualitative distribution and the change of the distribution does not depend on the orientation.
The exponential distribution shows the best agreement to the DDD data.
Thus, it can be concluded that an exponential distribution, e.g. as proposed by Sills et al. \cite{Sills_2018}, is a good approximation for the link length distribution. 
\begin{figure*}[t]
    \begin{subfigure}[b]{0.28\textwidth}
        \centering
        \captionsetup{justification=centering,margin=1.2cm}
        \includegraphics[height=4.5cm]{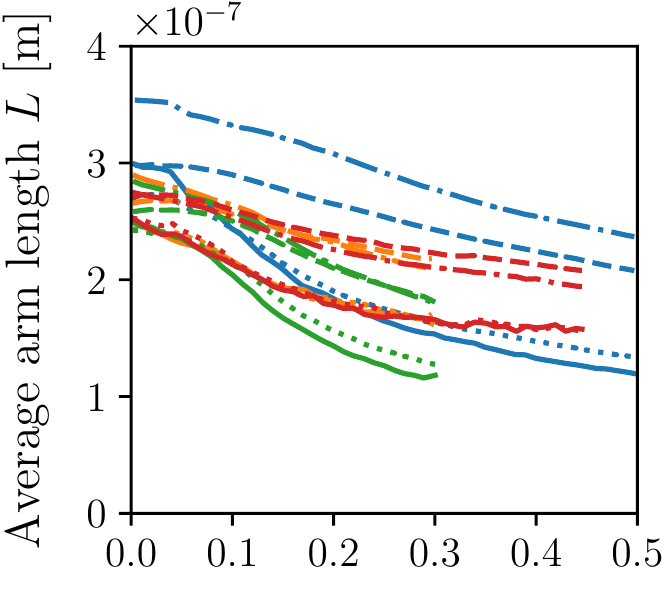}\llap{
          \parbox[b]{0.8cm}{(a)\\\rule{0ex}{4.15cm}
          }}
    \end{subfigure}
    \hspace*{-2mm}
    \begin{subfigure}[b]{0.1\textwidth}
        \centering
        \includegraphics[height=4.5cm]{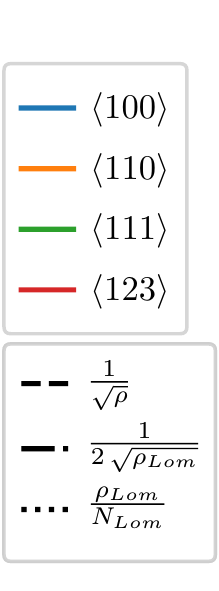}
    \end{subfigure}
    \hspace{1mm}
    \begin{subfigure}[b]{0.25\textwidth}
        \centering
        \includegraphics[height=4.5cm]{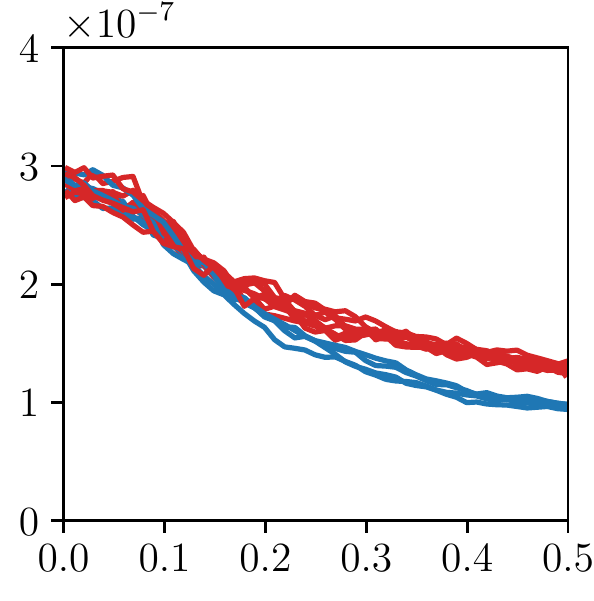}\llap{
          \parbox[b]{0.8cm}{(b)\\\rule{0ex}{4.15cm}
          }}\llap{
         \parbox[b]{2.5cm}{\large$\langle 100 \rangle$\\\rule{0ex}{3.6cm}
         }}
         \llap{\shortstack{%
         \includegraphics[scale=.65]{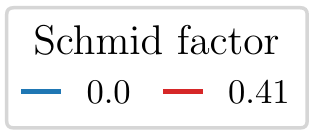}\\
         \rule{0ex}{2.6cm}%
         }
         \rule{0.3cm}{0ex}}
    \end{subfigure}
    \begin{subfigure}[b]{0.25\textwidth}
        \centering
        \includegraphics[height=4.5cm]{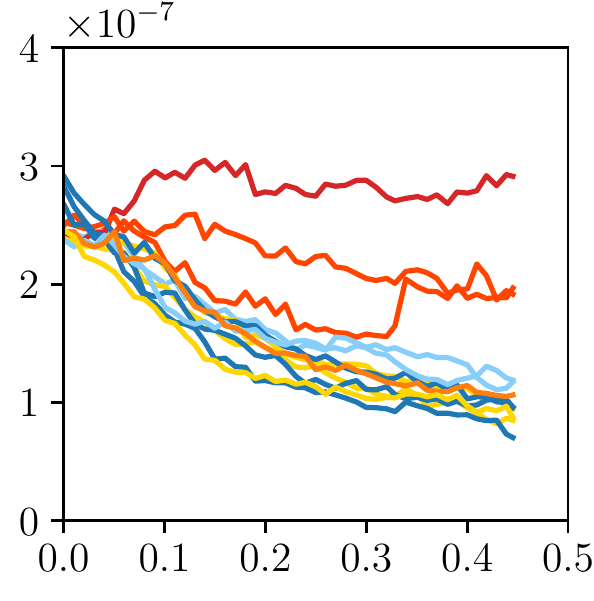}\llap{
          \parbox[b]{0.8cm}{(c)\\\rule{0ex}{4.15cm}
          }}\llap{
        \parbox[b]{2.5cm}{\large$\langle 123 \rangle$\\\rule{0ex}{3.6cm}
         }}
    \end{subfigure}
    \hspace*{-2mm}
    \begin{subfigure}[b]{0.1\textwidth}
        \centering
        \includegraphics[height=4.5cm]{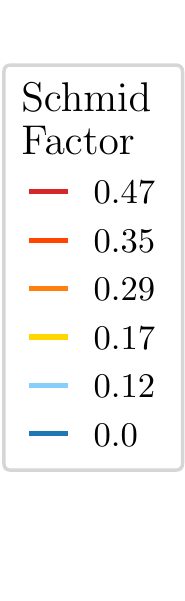}
    \end{subfigure}

    \begin{subfigure}[b]{0.28\textwidth}
        \centering
        \includegraphics[height=5cm]{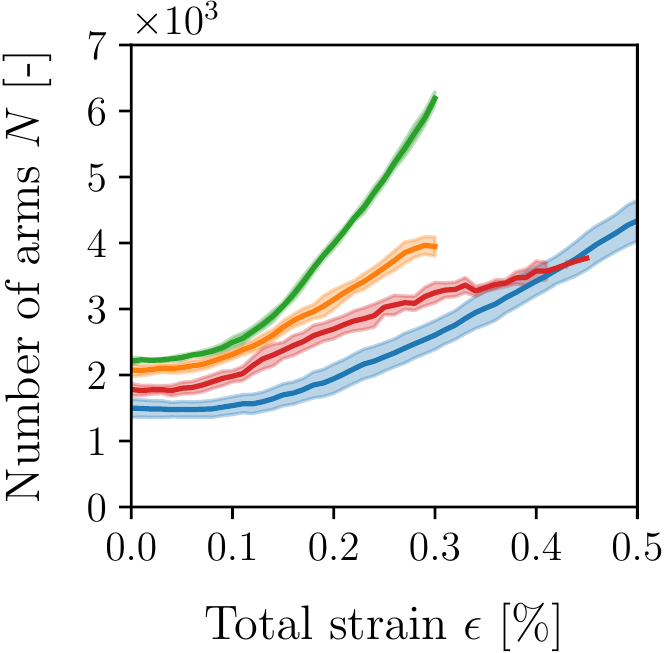}\llap{
          \parbox[b]{0.8cm}{(d)\\\rule{0ex}{4.65cm}
          }}
    \end{subfigure}
    \hspace*{-2mm}
    \begin{subfigure}[b]{0.10\textwidth}
        \centering
        \includegraphics[height=5cm]{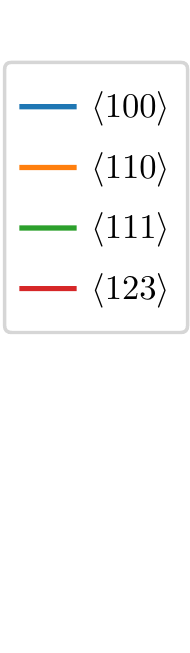}
    \end{subfigure}
    \hspace{1mm}
    \begin{subfigure}[b]{0.25\textwidth}
        \centering
        \includegraphics[height=5cm]{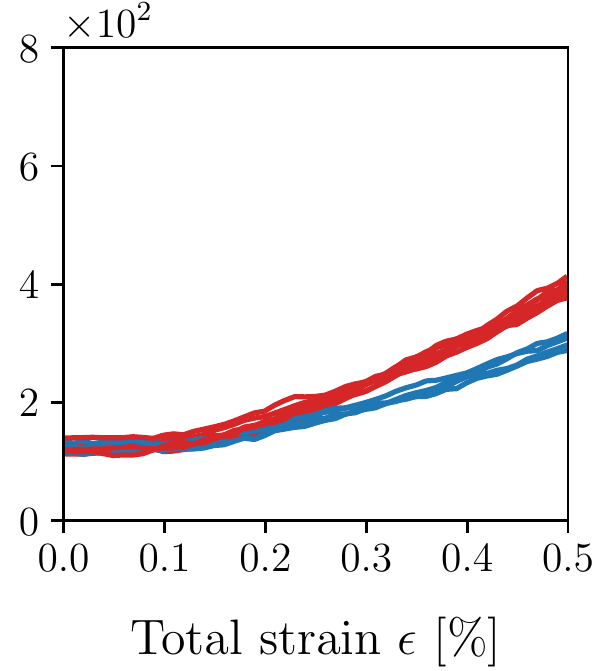}\llap{
          \parbox[b]{0.8cm}{(e)\\\rule{0ex}{4.65cm}
          }}\llap{
         \parbox[b]{2.5cm}{\large$\langle 100 \rangle$\\\rule{0ex}{4.1cm}
         }}
         \llap{\shortstack{%
         \includegraphics[scale=.65]%{plot_Lmean_eps_per_schmid_factor_100_leg.pdf}\\
         {plot_Lmean_schmid_100_leg.pdf}\\
         \rule{0ex}{3.1cm}%
         }\rule{0.3cm}{0ex}
         }
    \end{subfigure}
    \begin{subfigure}[b]{0.25\textwidth}
        \centering
        \includegraphics[height=5cm]{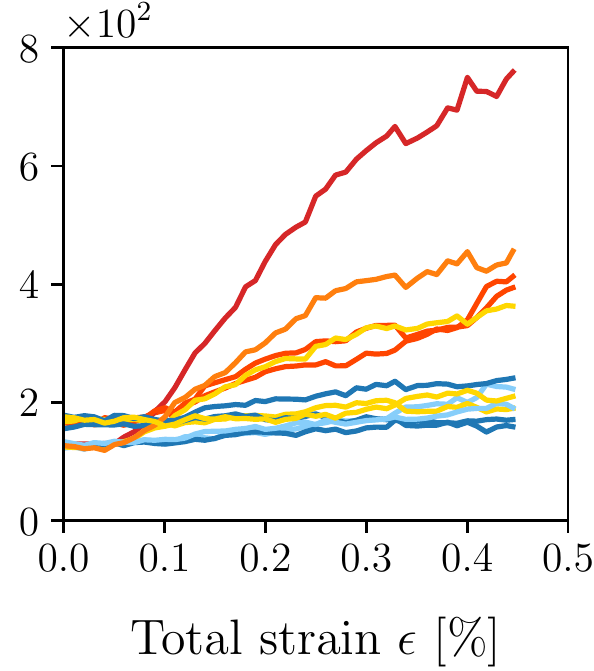}\llap{
          \parbox[b]{0.8cm}{(f)\\\rule{0ex}{4.65cm}
          }}\llap{
         \parbox[b]{2.5cm}{\large$\langle 123 \rangle$\\\rule{0ex}{4.1cm}
         }}
    \end{subfigure}
    \hspace*{-2mm}
    \begin{subfigure}[b]{0.1\textwidth}
        \centering
        \includegraphics[height=5cm]{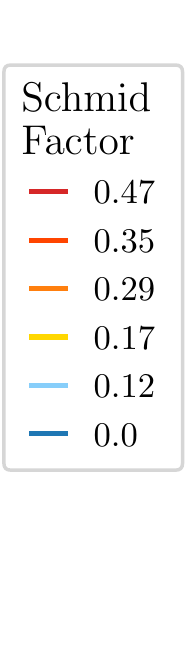}
    \end{subfigure}
    \caption{
    Comparison of the average link length (a) and number of links (d) between all investigated orientations. Slip system activity dependent comparison of the average link length in $\langle 100 \rangle$ orientation (b) and in $\langle 123 \rangle$ orientation (c) as well as slip system activity dependent comparison of the number of links in $\langle 100 \rangle$ orientation (e) and in $\langle 123 \rangle$ orientation (f). 
    }
    \label{fig:Schmid_Lomer_100_123}
\end{figure*}
\begin{figure*}[t]
    \centering
        \begin{subfigure}[t]{\textwidth}
        \centering
        \includegraphics[scale=0.8]{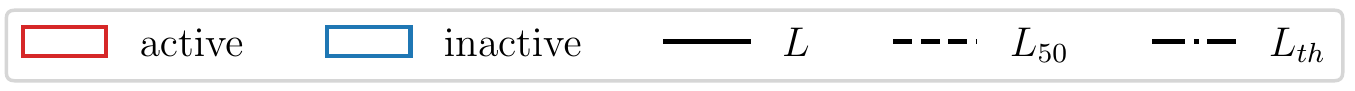}
    \end{subfigure}
    \begin{subfigure}[t]{0.26\textwidth}
        \centering
        \includegraphics[height=4.7cm]{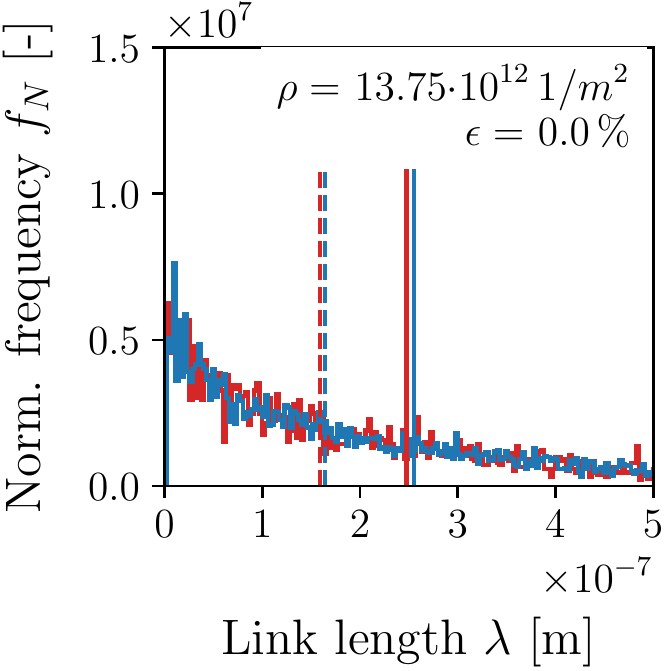}\llap{
          \parbox[b]{0.6cm}{(a)\\\rule{0ex}{4.4cm}
          }}
    \end{subfigure}
    \hfill
    \begin{subfigure}[t]{0.23\textwidth}
        \centering
        \includegraphics[height=4.7cm]{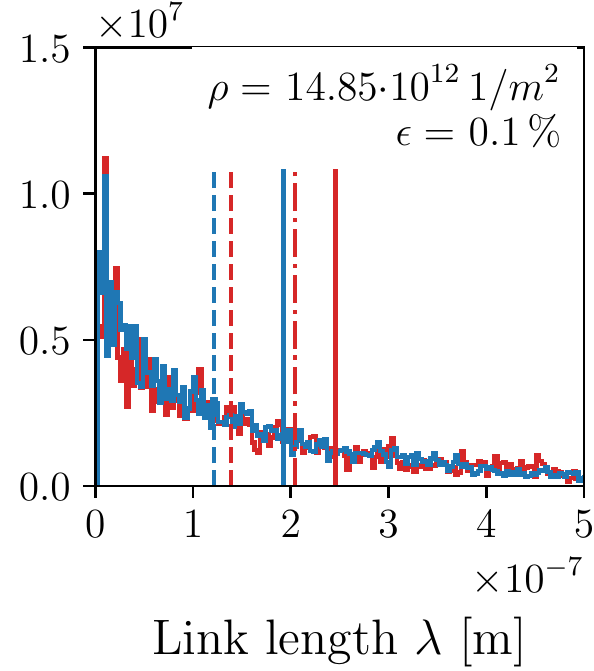}\llap{
          \parbox[b]{0.6cm}{(b)\\\rule{0ex}{4.4cm}
          }}
    \end{subfigure}
    \hfill
    \begin{subfigure}[t]{0.23\textwidth}
        \centering
        \includegraphics[height=4.7cm]{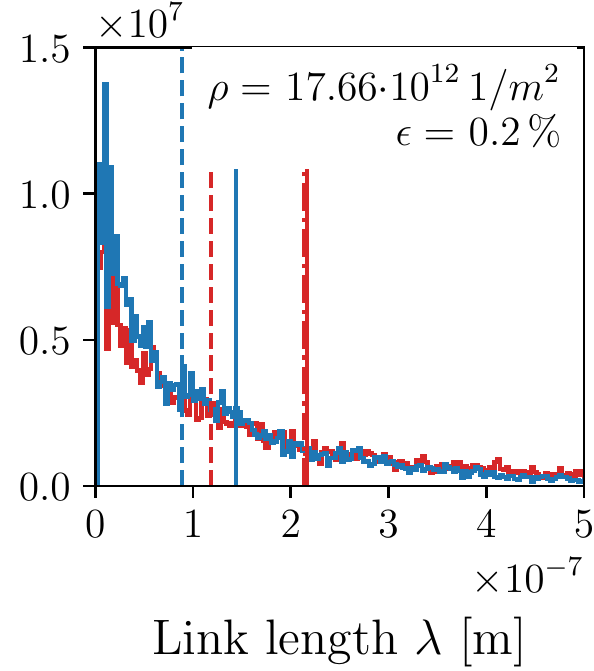}\llap{
          \parbox[b]{0.6cm}{(c)\\\rule{0ex}{4.4cm}
          }}
    \end{subfigure}
    \hfill
    \begin{subfigure}[t]{0.23\textwidth}
        \centering
        \includegraphics[height=4.7cm]{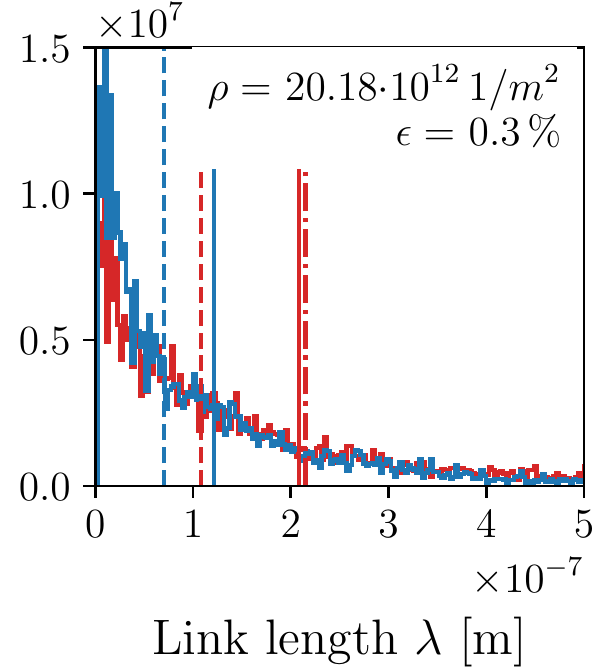}\llap{
          \parbox[b]{0.6cm}{(d)\\\rule{0ex}{4.4cm}
          }}
    \end{subfigure}
    \hfill
    \caption{Normalized frequency over link length in $\langle 123 \rangle$ orientation for (a) $0.0\%$, (b) $0.1\%$, (c) $0.2\%$ and (d) $0.3\%$ total strain. The average link length $L$ is indicated by solid, the median length $L_{50}$ by dashed and the threshold length $L_{th}$ by dash-dotted vertical lines. }
    \label{fig:freq_arm_123}
\end{figure*}

The exponential distribution uses as unique parameter the average link length $L$.
The average link length is shown in \autoref{fig:Schmid_Lomer_100_123}(a) for the investigated crystal orientations.
A common first approximation might be the consideration of the averaged dislocation spacing given by the $1/\sqrt{\rho}$ relation, which is shown by dashed lines in the diagram.
The averaged Lomer junction spacing is given by the $1/\sqrt{\rho_{Lom}}$ relation.
The Lomer density is derived on the basis of the total lengths of the Lomer junctions. 
Assuming that there are two Lomer arms between Lomer junctions, a factor of $0.5$ is chosen for the Lomer density approximation, which leads to $1/(2\sqrt{\rho_{Lom}})$ and is shown by the dash-dotted lines.
Additionally, the average Lomer arm length approximation with respect to the number of Lomer junctions $\rho_{Lom}/N_{Lom}$ as proposed by \cite{Sills_2018} is shown by dotted lines.

We observe a similar behavior for both square root approximations, with a slight shift for the absolute value of the approximated average link length compared to the ground truth data (see \autoref{fig:Schmid_Lomer_100_123}(a)).
We assume that this shift arises due to the accumulation
of further dislocation reactions between the Lomer arms in the dislocation network, resulting in shorter link lengths. 
The approximation $\rho_{Lom}/N_{Lom}$ shows a good fit, which indicates a correlation between the average length of the Lomer arm with the average length of the Lomer junction.
This is a surprising observation which can not be fully explained by the present data.

So far, the analysis of the link lengths has been considered for the entire system but not depending on the slip system.
However, the resolved shear stress differs for each slip system and influences the bow-out of dislocations.
As proposed in~\cite{2022_Katzer_JotMaPoS}, we apply a Schmid factor (orientation) dependent approach for the classification of DDD data.
In the following, we examine the influence of the slip system activity on the link length characteristics.
We observe a difference for the evolution of the links between active and inactive slip systems.
The Schmid factor dependent results of the average link length and the number of links are shown in \autoref{fig:Schmid_Lomer_100_123}, where (b) and (e) show the results for the high symmetry $\langle 100 \rangle$ orientation and (c) and (f) for the $\langle 123 \rangle$ orientation, respectively.
For comparison, (d) shows the evolution of the number of links for each orientation.
We observe, that the average link length starts to deviate between active and inactive slip systems at $0.1\%$~total strain with an increase of the deviations for larger plastic deformation.
It can be observed that the average link length on active slip systems decreases less compared to inactive systems.

In $\langle 123 \rangle$ orientation, we even observe a slight initial increase of the average link length of the most active slip system.
The reduction of the average link length on inactive slip systems is in accordance with our expectation of a densifying dislocation network.
When looking at the number of links, it is observed that it evolves differently for each orientation (cp. \autoref{fig:Schmid_Lomer_100_123}(d)).
With respect to the slip system activity, the number of links increase stronger on active than on inactive slip systems with increasing plastic deformation.
This effect is more pronounced for the most active slip system in $\langle 123 \rangle$ orientation.
Thus, the slip system activity has an influence on the formation and stability of the Lomer arms.

The influence of the slip system activity on the distribution of the link length and its evolution is shown in~\autoref{fig:freq_arm_123} for the $\langle 123 \rangle$ orientation for a total strain from $0.0\%$ to $0.3\%$.
We use the distinction of slip system activity in $\langle 123 \rangle$ orientation with a threshold value for the Schmid factor of 0.25, leading to four active slip systems. 
At a total strain of $0.1\%$, the link length distribution between active and inactive is similar.
With increasing plastic deformation, we observe small deviation of the distributions, also visible by the corresponding median and average link lengths.
The distributions on inactive slip systems shifts stronger to shorter link lengths compared to the active slip systems during plastic deformation.
The average link length decreases for inactive as well as for active slip systems, however, the reduction is more pronounced for inactive slip systems.

\begin{figure*}[b]
    \begin{subfigure}[t]{0.25\textwidth}
        \centering
        \includegraphics[height=4.7cm]{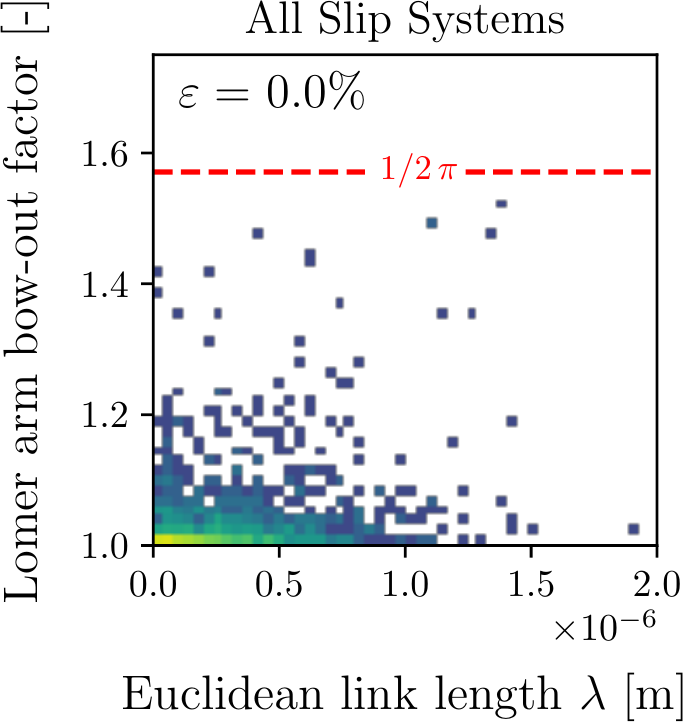}\llap{
          \parbox[b]{3.9cm}{(a)\\\rule{0ex}{4.4cm}
          }}
    \end{subfigure}
    \hfill
    \begin{subfigure}[t]{0.22\textwidth}
        \centering
        \includegraphics[height=4.7cm]{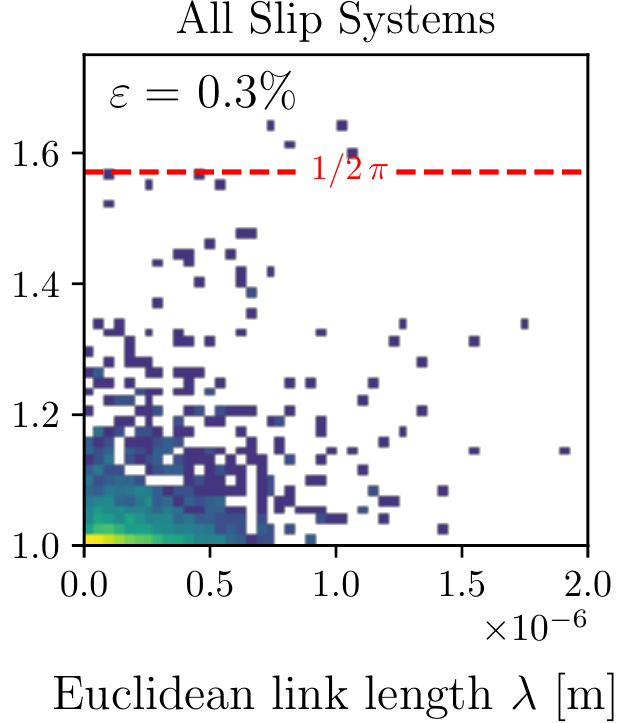}\llap{
          \parbox[b]{3.9cm}{(b)\\\rule{0ex}{4.4cm}
          }}
    \end{subfigure}
    \hfill
    \begin{subfigure}[t]{0.22\textwidth}
        \centering
        \includegraphics[height=4.7cm]{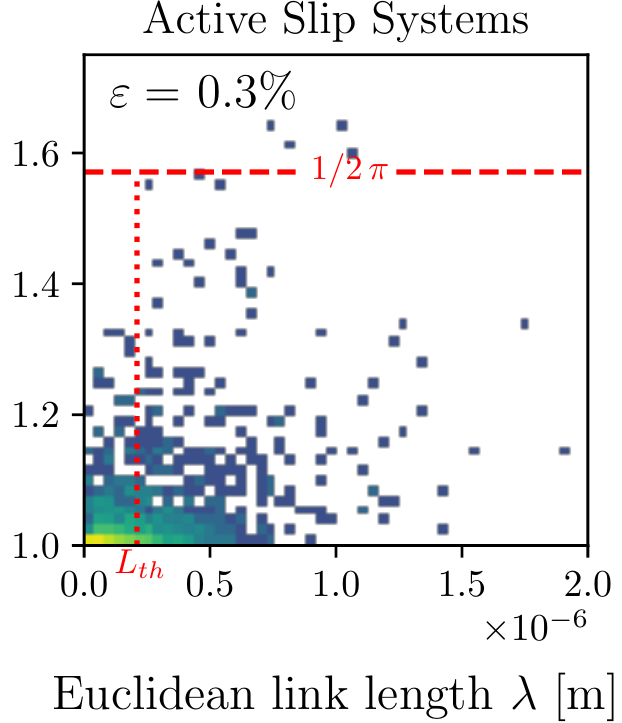}\llap{
          \parbox[b]{3.9cm}{(c)\\\rule{0ex}{4.4cm}
          }}
    \end{subfigure}
    \hfill
    \begin{subfigure}[t]{0.28\textwidth}
        \centering
        \includegraphics[height=4.7cm]{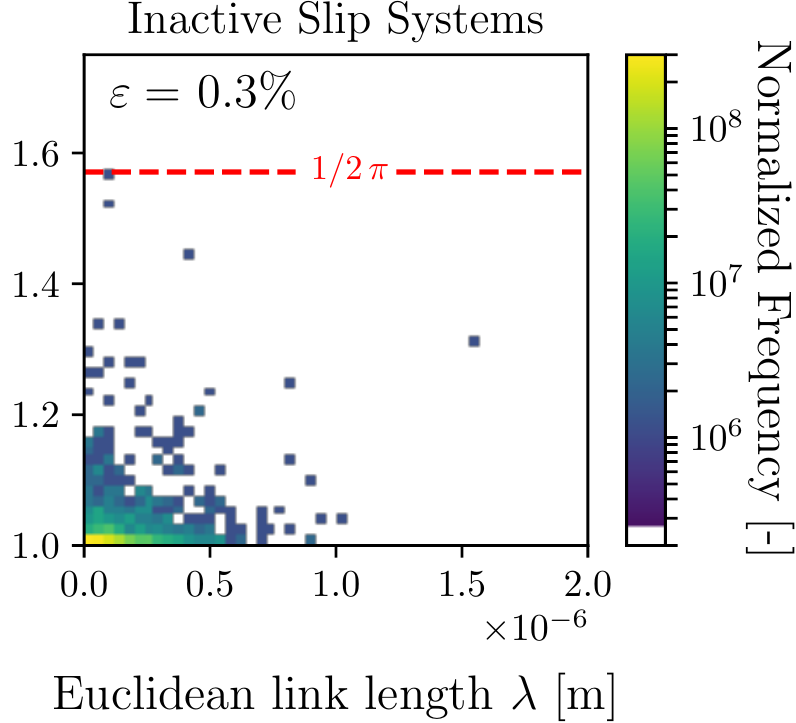}\llap{
          \parbox[b]{5.0cm}{(d)\\\rule{0ex}{4.4cm}
          }}
    \end{subfigure}
    \caption{
    Lomer arm bow-out factor versus Euclidean link length at (a) the initial state and at (b) $0.3\%$ total strain as well as the separate evaluation of the simulation at a total strain of $0.3\%$ for (c) active slip systems and (d) inactive slip systems for one sample with $\langle 123 \rangle$ orientation. The color indicates the normalized frequency. The red dashed horizontal line indicates a semi-circle bow-out, which corresponds to a bow-out factor $\pi/2$. The red dotted vertical line in (c) indicates the threshold length $L_{\mathrm{th}}$.
    }
    \label{fig:Bow-out-length}
\end{figure*}

The physical origin of the slightly different link length evolution of the active resp. inactive slip systems is addressed here: 
Active slip systems show less decrease in link length due to higher resolved shear stresses compared to inactive slip systems.
A higher resolved shear stress leads to bow-out and thus lengthening of the links, while densification of the network leads to an overall decrease of the link length.
The bow-out depends on the magnitude of the resolved shear stress.
A Lomer arm bow-out factor is calculated by dividing the link length with the Euclidean distance between the start and end-node of the Lomer arm, here referred to as ''Euclidean link length''.
\autoref{fig:Bow-out-length} shows the probability distribution of the Euclidean link length as a function of its bow-out.
Compared to the initial state (\autoref{fig:Bow-out-length}(a)), the evolved network (\autoref{fig:Bow-out-length}(b)) exhibits more curved links.
However, many links are short and straight at both strains.
By classification of the evolved network into active (\autoref{fig:Bow-out-length}(c)) and inactive (\autoref{fig:Bow-out-length}(d)) slip systems, we can relate long and strongly curved links to active slip systems.
However, we observe also bow-out of the links on inactive slip systems.
The length and the bow-out of a link is limited due to the increasing probability of re-reaction with other dislocations and due to unzipping of Lomer junctions.

To classify the length and the bow-out of the links, the results are compared with a line tension model.
The line tension model is used to calculate the threshold stress for the link instability.
Several line tension models exist in the literature, e.g. \cite{saada_sur_1960,Schoeck1972,Baird1965,Foreman1967}.
The models describe the relation between the Euclidean link length and the resolved shear stress. 

Since most of the links bow-out much less than a semi-circle, the prefactor of the line tension model is assumed to be reduced \cite{Foreman1967}.
This is why in this work, a simplified approach with a small prefactor from Rodney et al. \cite{Rodney1999} for the calculation of the threshold $L_{th}$ of the Euclidean link length is used, which is calculated by
\begin{equation}\label{eq:L_th}
    L_{th} \approx 0.5 \mu b \frac{1}{\left| \tau_{\mathrm{eff}} \right|}.
\end{equation}
Here, b is the Burgers vector magnitude, $\mu$ is the shear modulus and $\tau_{\mathrm{eff}}$ is the resolved shear stress on a slip system.
The dash-dotted lines in \autoref{fig:freq_arm_123} as well as the dotted line in \autoref{fig:Bow-out-length}(c) indicate the threshold link lengths.
We observe link lengths longer than the threshold link length.
Thus, parts of the Lomer junctions remain stable on active slip systems even if the shear stresses, calculated with the externally applied load only, exceed the critical shear stress of the Lomer arms.
This finding suggests that the Lomer junction stability in a dislocation network cannot be replaced by a general analytic term without further network information, e.g. internal stress contributions.

Hence, the use of a simplified stability threshold value to define a maximum link length as used in some averaged descriptions in the literature, e.g.\cite{Rodney1999,Sudmanns_2020}, has to be reconsidered.
We are aware that more sophisticated line tension models approximate the threshold stress more precisely, e.g. \cite{Hirth1966,Foreman1967}, containing additional length and angle dependent terms, which hinders the calculation of an unambiguous threshold length solution.
However, independent of the applied line tension model, we assume that the stability of these long links could be favored by the increasing presence of other reactions and internal stresses. 
This needs further investigation.

Summarizing the findings of this paper, we investigated the Lomer arm length distribution in dislocation networks at different total strains
and showed that the distribution is well described by an exponential distribution.
The crystal orientation has been found to have barely any influence on the distribution 
of the Lomer arm lengths obtained from the overall structure.
The modeling of the average Lomer arm length based on the Lomer junction density showed good results and provides an alternative to the general formulation.
If the network information exists, an approximation by $\rho_{Lom}/N_{Lom}$ can be applied.
However, the number of Lomer junctions might be not accessible for many continuum frameworks.
In addition, it has to be remarked, that the number of Lomer junction is found to correlate with the strain rate \cite{Fan2021}.
This correlation has not been part of this study but needs further investigation.

Furthermore, we showed that the Lomer arm length distribution and the average Lomer arm length differ between active and inactive slip systems.
Lomer arms on active slip systems show longer arm lengths compared to inactive slip systems.
The analysis of the Lomer arm bow-out showed that the bow-out on active slip systems is larger compared to inactive slip systems. However, bow-outs can also be observed in inactive slip systems.
An investigation of a theoretical maximum Lomer arm length and the observation of Lomer arms that exceed the threshold value of~\autoref{eq:L_th} lead to the conclusion that some of the Lomer junctions persist in the network even with long Lomer arms.
In current continuum modeling of the junction stability, the presence of such large arms in stable structures is not considered \cite{Sudmanns_2020}.
Therefore, future studies will analyze the Lomer junction dissolution process within a dislocation network dynamically.\\

%TC:ignore
\noindent\textbf{Declaration of Competing Interest} 

The authors declare that they have no known competing financial interests or personal relationships that could have appeared to influence the work reported in this paper.\\

\noindent\textbf{Acknowledgements}

The financial support for this work in the context of the DFG research project SCHU 3074/4-1 is gratefully acknowledged.
This work was performed on the computational resource HoreKa funded by the Ministry of Science, Research and the Arts Baden-Württemberg and DFG.
%TC:endignore

%\input{Acknowledgements.tex}

%\section{Introduction}

%\section*{Acknowledgments}



%\bibliographystyle{elsarticle-harv}
%\bibliographystyle{elsarticle-num.bst}
\bibliographystyle{model1a-num-names-test.bst} % Scripta

%Change Spacing between bibfiles
\let\OLDthebibliography\thebibliography
\renewcommand\thebibliography[1]{
  \OLDthebibliography{#1}
  \setlength{\parskip}{0pt}
  \setlength{\itemsep}{0pt plus 0.3ex}
}


%\newpage
%\section*{References}
%\zkfrage{Welche Zitierstil? DOI? Reihenfolge? (Alphabetisch? Chronologisch?)Url nur hinterlegt, wenn man auf die Referenz klickt?}
%\bibliographystyle{elsarticle-num}
%\bibliography{FOR1650}
%\bibliography{multiplication_Literature}
{\footnotesize
\bibliography{DataAnalysis}
}



% \end{thebibliography}

\end{document}